\documentclass[10pt,a4paper]{article}
\usepackage{amsmath}
\usepackage{amssymb}
\usepackage{amsfonts}
\usepackage{amstext}
\usepackage{amsbsy}
\usepackage[mathscr]{eucal}
\usepackage{graphicx}
\usepackage{color}
\usepackage[all]{xy}
\newcommand{\beq}{\begin{equation}}
\newcommand{\eeq}{\end{equation}}
\newcommand{\beqa}{\begin{eqnarray}}
\newcommand{\eeqa}{\end{eqnarray}}
\newcommand{\ket}[1]{| #1 \rangle}

\newtheorem{thm}{Theorem}[subsection]

 \numberwithin{equation}{subsection}

\makeindex
\title{\Large\textbf{Generalized controlled phase QGE}}

\author{\textit{ Hoshang Heydari}\\
        \small\textit{Department of Physics, Stockholm university 10691 Stockholm Sweden}\\
\\\small\textit{Email: hoshang@physto.se}}

\date{}
\begin{document}

\maketitle \thispagestyle{empty}

\begin{abstract}
We construct a generalized controlled phased gate entangler for a
multi-qubit state based on the geometrical structure of  quantum
systems. We also investigate relation between the generalized
controlled phase construction of a quantum gate entangler and graph
state for two-qubit and three-qubit states.
\end{abstract}
 \maketitle
\section{Introduction}
 There are several schemes for quantum computation, e.g., standard
scheme which is called  quantum circuit model. An alternate scheme
is called measurement-based quantum computation which is also called
cluster state scheme or one-way quantum computer. The structure of
cluster state scheme is very interesting and it is not yet fully
uncovered \cite{Raus,Nil,Hein}.  In the standard scheme the
processing of the qubit is based on a sequence of one-qubit and
two-qubit quantum gates which are unitary transformations. However,
cluster state model differs from unitary scheme of quantum
computation. One important class of these quantum gates is called
the quantum gate entangler such as controlled not or controlled
phase gates.
 Recently, we  have constructed quantum gate entangler for multi-qubit states and
   general multipartite states \cite{Hosh1,Hosh2}. In this paper we will
   construct quantum gate entangler based on generalized controlled phase gate
    for multi-qubit states. In section \ref{SPRT} we will give a short
   introduction to the standard quantum controlled phased gate and construction of cluster state based on this
   quantum gate entangler.
   In section \ref{gen} we construct  a generalized controlled phase
   gate for multi-qubit state. Then we will show how this gate could
   transform any product state to an entangle state. In section \ref{three} we will
   give an illustrative example of generalized controlled phase
   gate construction for three-qubit states and we compare this
   construction with a three-qubit graph state. We also give a physical implementation of control phase gate
   and its generalization for multi-qubit states. This is a generalization of Hamiltonian of
   two spin systems. In following sections we need the definition of a
    complex projective space $\mathbb{CP}^{n-1}$ which is defined to be the
set of lines through the origin in $\mathbb{C}^{n}$, that is, $
\mathbb{CP}^{n-1}=\mathbb{C}^{n}-{0}/ \sim$, where the equivalence
relation $\sim$ is defined as follow;
$(x_{0},\ldots,x_{n})\sim(y_{0},\ldots,y_{n})$ for $\lambda\in
\mathbb{C}-0$, where $y_{i}=\lambda x_{i}$ for all $ 0\leq i\leq n$.

\section{Controlled phase gate and cluster state}\label{SPRT}
In this section we will give short introduction to controlled phase
gate and cluster state generated by this gate. Let $Z=\left(
                                                        \begin{array}{cc}
                                                          1 & 0 \\
                                                          0 &-1 \\
                                                        \end{array}
                                                      \right)$ be
                                                      a gate that
                                                      acts on a qubit as
                                                      $Z\ket{0}=\ket{0}$ and $Z\ket{1}=-\ket{1}$
                                                      Then,
 in a
controlled phase gate $CZ$, the first qubit acts as the control
qubit and the second qubit acts as the target qubit. If the first
qubit is 1, then we apply a $Z$ gate to the second qubit and if
first qubit is 0, then we apply an identity gate to the second qubit.
In matrix form it is given by
\begin{equation}
CZ=\frac{1}{2}(I_{2}\otimes I_{2}+I_{2}\otimes Z+Z\otimes
I_{2}-Z\otimes Z),
\end{equation}
where $I_{2}$ is a 2-by-2 identity matrix. This controlled phase
gate can entangle a quantum state. For example, let
$\ket{+}=\frac{1}{\sqrt{2}}(\ket{0}+\ket{1})$. Then, by applying the
$CZ$ to a product two-qubit state we get a cluster state
\begin{equation}
CZ\ket{+}\otimes\ket{+}=\frac{1}{2}(\ket{00}+\ket{01}+\ket{10}-\ket{11}).
\end{equation}
It is easy to check that this state is an entangled two-qubit state.
 Note that
the $CZ$ gate is a special case of the following general gate
\begin{equation}
\mathcal{Z}_{4\times 4}=\left(
                \begin{array}{c|c}
                 U_{0} & 0 \\
                  \hline
                  0 & U_{1} \\
                \end{array}
              \right),
\end{equation}
where $U_{0}=\left(
               \begin{array}{cc}
                 e^{i\varphi_{0}} & 0 \\
                 0 & e^{i\varphi_{1}} \\
               \end{array}
             \right)$, and $U_{1}=\left(
               \begin{array}{cc}
                 e^{i\varphi_{2}} & 0 \\
                 0 & e^{i\varphi_{3}} \\
               \end{array}
             \right)$, for
             $e^{i\varphi_{0}}=e^{i\varphi_{1}}=e^{i\varphi_{2}}=1$,
             and $e^{i\varphi_{3}}=-1$.
In following section we will generalize this construction of
controlled phase gate for a two-qubit state into a multi-qubit
state.

\section{Generalized controlled phase gate entangler for  multi-qubit states}\label{gen}
In this section we will construct  a generalized controlled phase
   gate entangler for multi-qubit state.  We will
also in detail discuss this operator for three-qubit states.
For a multi-qubit state
\begin{equation}\ket{\Psi}=\sum^{1}_{x_{m-1},x_{m-2},\ldots,
x_{0}=0}\alpha_{x_{m-1}x_{m-2}\cdots x_{0}}\ket{x_{m-1}x_{m-2}\cdots
x_{0}}
\end{equation}
 we define a
generalized controlled phase gate $\mathcal{Z}_{2^{m}\times 2^{m}}$
by
\begin{eqnarray}\label{Rmat}
\mathcal{Z}_{2^{m}\times2^{m}}&=&\mathcal{Z}^{0}_{2^{m}\times2^{m}}\mathcal{Z}^{1}_{2^{m}\times2^{m}}
\cdots\mathcal{Z}^{2^{m-1}-1}_{2^{m}\times2^{m}}\\&=& \left(
  \begin{array}{cccc}
    U_{0} & 0 & \cdots & 0 \\
    0 & U_{1} & & \vdots \\
    \vdots & & \ddots & 0 \\
    0 & \cdots & 0 &U_{2^{m-1}-1} \\
  \end{array}
\right)\\\nonumber&=&\bigoplus^{2^{m-1}-1}_{x=0} U_{x},
\end{eqnarray}
where $U_{x}=\left(
                                                 \begin{array}{cc}
                                                   \alpha_{x} & 0 \\
                                                   0 & \alpha_{x+1} \\
                                                 \end{array}
                                               \right)$ for all $x=x_{m-1}2^{m-1}+x_{m-2}2^{m-2}+\cdots
+x_{0}2^{0}$. For example for $x=0$ we have
\begin{equation}
\mathcal{Z}^{0}_{2^{m}\times2^{m}}=\left(
                \begin{array}{c|c}
                 U_{0} & 0_{2\times2^{m}-2} \\
                  \hline
                  0_{2^{m}-2\times2} & I_{2^{m}-2\times2^{m}-2} \\
                \end{array}
              \right),
\end{equation}
where $0_{2\times2^{m}-2}$ and $ 0_{2^{m}-2\times2}$ are zero
matrices  and $I_{2^{m}-2\times2^{m}-2}$ is an identity matrix. Note
that  all  decomposition matrices
$\mathcal{Z}^{i}_{2^{m}\times2^{m}}$ and
$\mathcal{Z}^{j}_{2^{m}\times2^{m}}$ for all $i, j=0,1,\ldots, m-1$
commute, that is
$[\mathcal{Z}^{i}_{2^{m}\times2^{m}},\mathcal{Z}^{j}_{2^{m}\times2^{m}}]=0$.

There is also a geometrical interpretation of this transformation
based on the Lie group. We define a Lie group to be a differentiable
manifold $G$ that satisfies a product map $G\times G\longrightarrow
G$ given by $(g,h)\longmapsto gh$ which make $G$ into a group. This
map and the map $G\longrightarrow G$ given by $g\longmapsto g^{-1}$
need to be differentiable. An example of Lie group $G=T^{2^{m}}$
is the abelian group of diagonal matrices of the form
\begin{equation}
A=\mathrm{Diag}(e^{i\varphi_{0}},
e^{i\varphi_{0}},\ldots,e^{i\varphi_{2^{m}}})\in G.
\end{equation}
This Lie group is topologically $S^{1}\times S^{1}\times\cdots\times
S^{1}$, that is $2^{m}$-times topological product of $S^{1}$ which
can be identified by $2^{m}$-torus. The subgroup $T^{2^{m}}$ of unitary
group $U(2^{m})$ is called a maximal torus of $U(2^{m})$ and any
conjugate  $hT^{2^{m}}h^{-1}$ of maximal torus is also called a maximal
torus. Thus, we can see that our transformation
$\mathcal{Z}_{2^{m}\times2^{m}}$ is a maximal torus if we assume
that
\begin{eqnarray}\nonumber
\alpha_{x_{m-1}x_{m-2}\cdots x_{0}}&=&|\alpha_{x_{m-1}x_{m-2}\cdots
x_{0}}|e^{i\varphi_{x_{m-1}x_{m-2}\cdots
x_{0}}}\\&=&e^{i\varphi_{x_{m-1}x_{m-2}\cdots x_{0}}}\in\mathbb{C}.
\end{eqnarray}
Then, we have following theorem.
\begin{thm} Let $\mathcal{Z}_{2^{m}\times
2^{m}}$ be a  generalized controlled phase gate given by equation
(\ref{Rmat}). Then, the state
\begin{equation}\mathcal{Z}_{2^{m}\times
2^{m}}(\ket{+}\otimes\ket{+}\otimes\cdots\otimes\ket{+})\in
\mathbb{CP}^{2^{m}-1} ,\end{equation} with
$\ket{+}=\frac{1}{\sqrt{2}}(\ket{0}+\ket{1})$ is entangled if and
only if elements of $\mathcal{Z}_{2^{m}\times 2^{m}}$ satisfy
\begin{eqnarray}\label{segreply1}
 &&\alpha_{x_{m-1}x_{m-2}\cdots x_{0}}\alpha_{y_{m-1}y_{m-2}\cdots
y_{0}}\\\nonumber&&\neq \alpha_{x_{m-1}x_{m-2}\ldots y_{j}\cdots
x_{0}}\alpha_{y_{m-1}y_{m-2} \cdots  x_{j}\ldots y_{0}}
\end{eqnarray}
\end{thm}
for all $j=0,1,\ldots,m-1$, where $\in \mathbb{CP}^{2^{m}-1}$ is
complex  projective space of dimension $2^{m}-1$.
The proof of this theorem follows from the Segre variety
\begin{eqnarray}\label{eq: submeasure}
&&\bigcap_{\forall j}\mathcal{V}(\alpha_{x_{m-1}x_{m-2}\cdots
x_{0}}\alpha_{y_{m-1}y_{m-2}\cdots y_{0}}\\\nonumber&&-
\alpha_{x_{m-1}x_{m-2}\ldots y_{j}\cdots
x_{0}}\alpha_{y_{m-1}y_{m-2} \cdots x_{j}\ldots
y_{0}})\\\nonumber&&=\mathrm{Im}\left(\overbrace{\mathbb{CP}^{1}\times\mathbb{CP}^{1}\times\cdots\times\mathbb{CP}^{1}}^{m
~\text{times}} \longrightarrow\mathbb{CP}^{2^{m}-1}\right)
\end{eqnarray}
which is the image of the Segre embedding and is an intersection of
families of quadric hypersurfaces in $\mathbb{CP}^{2^{m}-1}$ that
represents the completely decomposable tensor \cite{Hosh3}.


\section{Examples of controlled phase gate for multi-qubit systems}\label{three}
 To visualize our construction we will first  construct a generalized controlled phased gate entangler for
 a three-qubit state and then discuss a physically implementable construction of controlled phase gate for two-spin system  that can be
 generalized to multi-qubit systems. For a three-qubit system
$\mathcal{Z}_{8\times 8}$ is given by
\begin{equation}
\mathcal{Z}_{8\times 8}=\mathcal{Z}^{0}_{8\times
8}\mathcal{Z}^{1}_{8\times 8}\mathcal{Z}^{2}_{8\times
8}\mathcal{Z}^{3}_{8\times 8},
\end{equation}
where $\mathcal{Z}^{0}_{8\times 8}, \mathcal{Z}^{1}_{8\times 8},
\mathcal{Z}^{2}_{8\times 8}$, and $\mathcal{Z}^{3}_{8\times 8}$ are
defined as follows
\begin{equation}
\mathcal{Z}^{0}_{8\times 8}=\left(
  \begin{array}{c|c}
    U_{0}
   & 0 \\
    \hline
    0 & I_{6} \\
  \end{array}
\right),~ U_{0}=\left(
  \begin{array}{cc}
     e^{i\varphi_{000}}& 0 \\
    0 &  e^{i\varphi_{001}}\\
  \end{array}
\right),
\end{equation}
\begin{equation}
\mathcal{Z}^{1}_{8\times 8}=\left(
  \begin{array}{c|c|c}
    I_{2} & 0 &0\\
    \hline
    0 & U_{1}&0 \\
    \hline
    0 & 0&I_{4} \\
  \end{array}
\right)~ U_{1}=\left(
  \begin{array}{cc}
     e^{i\varphi_{010}}& 0 \\
    0 &  e^{i\varphi_{011}}\\
  \end{array}
\right),
\end{equation}
\begin{equation}
\mathcal{Z}^{2}_{8\times 8}=\left(
  \begin{array}{c|c|c}
  I_{4} & 0 &0\\
    \hline
    0 &U_{2} & 0 \\
    \hline
    0 & 0&I_{2} \\
  \end{array}
\right),  ~ U_{2}=\left(
  \begin{array}{cc}
     e^{i\varphi_{100}}& 0 \\
    0 &  e^{i\varphi_{101}}\\
  \end{array}
\right),~\text{and}
\end{equation}
\begin{equation}
\mathcal{Z}^{3}_{8\times 8}=\left(
  \begin{array}{c|c}
  I_{6}
 & 0 \\
    \hline
    0 &  U_{3}\\
  \end{array}
\right), ~U_{3}=\left(
  \begin{array}{cc}
     e^{i\varphi_{110}}& 0 \\
    0 &  e^{i\varphi_{111}}\\
  \end{array}
\right),
\end{equation}
As an important but special case let
$e^{i\varphi_{000}}=e^{i\varphi_{001}}=e^{i\varphi_{010}}=e^{i\varphi_{100}}=1$
and
$e^{i\varphi_{011}}=e^{i\varphi_{101}}=e^{i\varphi_{110}}=e^{i\varphi_{111}}=-1$.
Then, we have following cluster state
\begin{eqnarray}\nonumber
&&\mathcal{Z}_{8\times 8}(\ket{+}\otimes\ket{+}\otimes\ket{+})=
\frac{1}{\sqrt{8}}(\ket{000}+\ket{001}\\\nonumber&&+\ket{010}-\ket{011}+\ket{100}+\ket{101}-
\ket{110}-\ket{111}).
\end{eqnarray}
Note that this state is entangled since e.g.,
$\alpha_{000}\alpha_{110}\neq\alpha_{010}\alpha_{100}$. This is also
a graph state with three nodes connected which is obtained by
applying the controlled phased gate $CZ$ three times on node 1 and
2, 2 and 3, and 1 and 3.

Next, as an illustrative and physically implementable  example of
our construction we will discuss the construction of a two-qubit
gate \cite{Chen}. For example let the Hamiltonian of a two-spin
system without transverse field be given by
\begin{equation}
H=\frac{1}{2}\left(\sum^{2}_{i=1}\omega_{i}\sigma^{i}_{z}+J\sigma^{1}_{z}\sigma^{2}_{z}\right),
\end{equation}
where $\omega_{i}$ is the frequency of spin $i$, $J$ is the coupling
coefficient, and $\sigma^{i}_{z}$ is the $z$ projection operator of
spin $i$. Then, we have
\begin{eqnarray}
&&H(\ket{00}+\ket{01}+\ket{10}+\ket{11})\\\nonumber&&=\psi_{00}\ket{00}+\psi_{01}\ket{01}
+\psi_{10}\ket{10}+\psi_{11}\ket{11},
\end{eqnarray}
where $\psi_{00}=\frac{1}{2}(\omega_{1}+\omega_{2}+J)$,
$\psi_{01}=\frac{1}{2}(\omega_{1}-\omega_{2}-J)$,
$\psi_{10}=\frac{1}{2}(-\omega_{1}+\omega_{2}-J)$, and
$\psi_{11}=\frac{1}{2}(-\omega_{1}-\omega_{2}+J)$. Thus, the
evolution of this two-partite system is give by
\begin{eqnarray}
\ket{\psi(t)}&=&e^{-iHt}\ket{\psi(0)}\\\nonumber&=& \left(
     \begin{array}{cccc}
       e^{-i\psi_{00}t} & 0 & 0 & 0 \\
       0 & e^{-i\psi_{01}t} & 0 & 0 \\
       0 & 0 & e^{-i\psi_{10}t} & 0 \\
       0 & 0 & 0 & e^{-i\psi_{11}t} \\
     \end{array}
   \right)\ket{\psi(0)} .
\end{eqnarray}
Now, if we substitute $-\psi_{x_{1}x_{0}}t=\varphi_{x_{1}x_{0}}$,
for all $x_{1}, x_{0}=0,1$, then we recognize our
$\mathcal{Z}_{2\times 2}$ matrix for this two-partite system. This
construction can be generalized for multi-qubits systems, where the
Hamiltonian acting on a multi-qubit state as follows
\begin{eqnarray}\nonumber
&&H(\ket{00\cdots0}+\ket{00\cdots1}+\cdots+\ket{11\cdots1})\\\nonumber&&
=\psi_{00\cdots0}\ket{00\cdots0}+\psi_{00\cdots1}\ket{00\cdots1}\\&&+\cdots+\psi_{11\cdots1}\ket{11\cdots1},
\end{eqnarray}
where $\psi_{x_{m-1}x_{m-2}\cdots
x_{0}}=-\varphi_{x_{m-1}x_{m-2}\cdots x_{0}}/t$ describes a specific
multi-qubit system which we have explicitly defined for two-qubit
spin systems.

In this paper we have constructed a geometrical controlled phased
gate entangler for multi-qunit states based on a generalization of
controlled phase gate. The construction is simple and works for any
multi-qubit quantum states. In continuation of this work, it  would also
 be interesting to investigate the use of this method for
construction of generalized graph states which are building blocks of
measurement-based quantum computer models.

\end{document}